\documentclass[aps,pre,12pt]{revtex4}
\usepackage{amsmath}

\begin{document}

\title{Wake field generation and nonlinear evolution in a magnetized
electron-positron-ion plasma}
\author{P. K. Shukla \cite{a)}, G. Brodin, M. Marklund and L. Stenflo \cite{b)}}
\affiliation{Department of Physics, Ume\aa~University, 
SE-90187 Ume\aa,~Sweden}
\begin{abstract}
The nonlinear propagation of a circularly polarized electromagnetic (CPEM)
wave in a strongly magnetized electron-positron-ion plasma is investigated.
Two coupled equations describing the interaction between a high-frequency
CPEM wave and the low-frequency electrostatic wake field are derived. It is
found that the generation of the wake fields depends on the presence of the
ion species. The wake field generation in turn leads to de-acceleration and
frequency down conversion of the electromagnetic pulse.
\end{abstract}

\maketitle

\newpage

\section{Introduction}

Electron-positron (pair) plasmas are composed of electrons and positrons
which have the same mass and opposite charge. Such pair plasmas appeared in
the early Universe \cite{Misner73,Rees83}, and are frequently encountered in
bipolar outflows (jets) in active-galactic nuclei \cite{Begelman84,Miller87}%
, in micro-quasars \cite{Fender04}, in pulsar magnetospheres \cite
{Goldreich69,Michel82,Michel91,Beskin93}, in magnetars \cite{Marklund06}, in
cosmological gamma ray fireballs \cite{Piran04}, in solar flares \cite
{Hansen88}, and at the center of our galaxy \cite{Burns83}. Multiterawatt
and petawatt laser pulses interacting with solid density matters can create
pair plasmas \cite{Berezhiani92,Liang98,Gahn00,Wilks05}. Colliding
electromagnetic pulses also generate pairs from vacuum \cite{Narozhny2004}.
The pair plasmas at the surface of fast rotating neutron stars and magnetars
are held in strong magnetic fields, while super-strong magnetic fields can
be created in intense laser-plasma interaction experiments. Accordingly, the
understanding of collective phenomena in strongly magnetized pair plasmas
has been a topic of significant interest \cite
{Gedalin85,Shukla86,Iwamoto93,Brodin2007}. Specifically, it is to be noted
that in a magnetized pair plasma, we have new wave modes whose counterparts
do not exist in an electron-ion magnetoplasma \cite{Shukla84}. Furthermore,
both astrophysical and laboratory plasmas contain a fraction of ions besides
the electrons and positrons. In such environments, the linear and nonlinear
wave propagation characteristics are modified \cite
{Rizzato88,Berezhiani92a,Mahajan94,Mahajan95,Popel95,
Shukla2003,Kim2004,Cattert05,Marklund2005}.

In this paper, we consider the nonlinear propagation of magnetic
field-aligned circularly polarized electromagnetic (CPEM) waves in a
pair-ion magnetoplasma, to show the possibility of exciting electrostatic
plasma wakes by the ponderomotive force of the CPEM waves. Our manuscript is
organized in the following fashion. In Sec. II we discuss the linear
dispersion relation, the group velocity and the group dispersion for the
magnetic field-aligned CPEM waves in a pair-ion magnetoplasma. Section III
considers the nonlinear interactions between finite amplitude CPEM waves and
electrostatic plasma oscillations (EPOs), and presents the coupled
equations. Nonlinearities associated with the electron and positron mass
variations, as well as with the CPEM ponderomotive force are incorporated in
the dynamical equations for the modulated CPEM waves and the driven EPOs.
Section IV presents the conservation laws for nonlinearly coupled CPEM wave
EPO systems, and discusses the excitation of wakefields . Section V contains
a brief summary of our investigation.

\section{The linear dispersion relation}


We consider a magnetized quasi-neutral electron-positron-ion plasma, i.e. $%
n_{0e}=n_{0i}+n_{0p}$, where $n_{0e}$ is the background electron number
density, $n_{0p}$ is the background positron density, $n_{0i}$ is the
background ion density, and the background magnetic field is given by $\vec{B%
}_{0}=B_{0}\hat{z}$. For a right hand circularly polarized electromagnetic
(RCPEM) wave propagating along $\hat{z}$ with frequency $\omega \gg \omega
_{pi}=(4\pi n_{0i}e^{2}/m_{i})^{1/2}$, we have the dispersion relation \cite
{Stenflo1976,Shukla2006}

\begin{equation}
\frac{k^{2}c^{2}}{\omega ^{2}}=1-\frac{\omega _{pe}^{2}}{\omega (\omega
+\omega _{ce})}-\frac{\omega _{pp}^{2}}{\omega (\omega +\omega _{cp})},
\label{1}
\end{equation}
where $\omega _{pj}=(4\pi n_{0j}e^{2}/m_{e})^{1/2}$, with $j=e,p$, is the
plasma frequency, $\omega _{ce}=-eB_{0}/m_{e}=-\omega _{cp}$ is the electron
gyrofrequency, $e$ is the magnitude of the electron charge, and $m_{e}$ is
the electron mass (for left hand polarization $\omega +\omega _{c}$ is
replaced by $\omega -\omega _{c}$ \cite{Brodin1988}). In the limit of a
strong background magnetic field, i.e.\ $\omega \ll |\omega _{ce}|$, we can
simplify (\ref{1}) to obtain

\begin{equation}
\frac{k^{2}c^{2}}{\omega ^{2}}=1+\frac{\Omega _{i}^{2}}{\omega \omega _{cp}},
\label{eq:2}
\end{equation}
where we have introduced the notation $\Omega _{i}=(4\pi
n_{0i}e^{2}/m_{e})^{1/2}$ and used the background charge neutrality
condition. From Eq.\ (\ref{eq:2}) we obtain the group velocity

\begin{equation}
v_{g}=\frac{kc^{2}}{\omega +\Omega _{i}^{2}/2\omega _{cp}},
\label{eq:group-vel}
\end{equation}
and the group velocity dispersion $v_{g}^{\prime
}=dv_{g}/dk=(v_{g}/k)(1-v_{g}^{2}/c^{2})$. In the limit $kc\gg \omega
_{p}^{2}/\omega _{cp}$, we obtain

\begin{equation}
v_{g}\approx c\left[ 1-\frac{1}{2}\left( \frac{\Omega _{i}^{2}}{2\omega
_{cp}kc}\right) ^{2}\right] ,
\end{equation}
and $v_{g}^{\prime }\approx \Omega _{i}^{2}/4\omega _{cp}^{2}k^{3}c$, and we
have neglected the root with negative sign of the group velocity. Thus, in
this limit the RCPEM waves propagate with a speed close to the speed of
light.

\section{Nonlinear evolution equations}


Next, we consider a modulated nonlinear CPEM wave $(\omega ,k)$ , where the
ponderomotive force induces slowly ($\partial _{t}\ll \omega $) varying
electrostatic oscillations. We still assume that the motion is fast enough
for the ions to be immobile. The evolution equation for the vector potential
amplitude $A$ in the Coulomb gauge can then be written as \cite{Stenflo1985}

\begin{equation}
i\left( \partial _{t}+v_{g}\partial _{z}\right) A+\frac{1}{2}v_{g}^{\prime
}\partial _{z}^{2}A-\left[ \sum_{e,p}\frac{\omega \omega _{p}^{2}}{2(\omega
+\omega _{c})}\frac{v_{g}}{kc^{2}}\left( \frac{n_{1}}{n_{0}}\frac{\omega }{%
(\omega +\omega _{c})}-\frac{e^{2}\omega ^{3}|A|^{2}}{m_{e}^{2}c^{2}(\omega
+\omega _{c})^{3}}\right) \right] A=0,  \label{eq:vector-general}
\end{equation}
where the group velocity and dispersion are given above, and the indices e,p
have been omitted for notational convenience. $\ $The electrostatic
perturbations are described by the density fluctuation $n_{1}$, the
longitudinal velocity $v_{z}$ and the electrostatic potential $\phi $. These
variables satisfy the continuity equation

\begin{equation}
\frac{\partial n_{1}}{\partial t}+n_{0}\frac{\partial v_{z}}{\partial z}=0,
\label{eq:cont}
\end{equation}
and the momentum equation

\begin{equation}
m_{e}\frac{\partial v_{z}}{\partial t}=-qF-q\frac{\partial \phi }{\partial z}%
-\frac{T}{n_{0}}\frac{\partial n_{1}}{\partial z},
\end{equation}
where $q=-e$ for electrons and $q=e$ for positrons, together with

\begin{equation}
\frac{\partial ^{2}\phi }{\partial z^{2}}=4\pi e(n_{e1}-n_{p1}).
\label{eq:Poisson}
\end{equation}
Here $\ qF$ denotes the ponderomotive force, which can be written as \cite
{Stenflo1985}

\begin{equation}
qF=\frac{e^{2}\omega }{m_{e}(\omega +\omega _{c})}\left( \partial _{z}+\frac{%
k\omega _{c}}{\omega (\omega +\omega _{c})}\partial _{t}\right) |A|^{2},
\label{eq:ponderomotive}
\end{equation}
in a magnetized plasma. We assume that the temperature $T$ is the same for
electrons and positrons.

Next, we consider a high-frequency wave propagating with the group velocity,
such that the driven low-frequency perturbations to first order are
stationary functions of the comoving coordinate $z-v_{g}t$. Together with
the assumption $\omega \ll \omega _{c}$, the ponderomotive force (\ref
{eq:ponderomotive}) then reduces to

\begin{eqnarray}
qF &\simeq &\frac{e^{2}\omega }{m_{e}\omega _{c}}\left( 1-\sqrt{1+\frac{%
\Omega _{i}^{2}}{\omega \omega _{cp}}}\right) \partial _{z}|A|^{2}  \notag \\
&\simeq &-\frac{e^{2}\Omega _{i}^{2}}{2m_{e}\omega _{c}\omega _{cp}}\partial
_{z}|A|^{2},  \label{eq:pond-approx}
\end{eqnarray}
where we have used $\Omega _{i}^{2}/\omega \omega _{c}\ll 1$ in the last
step. \ Next we expand Eq. (\ref{eq:vector-general}) in the small parameter $%
\omega /\omega _{c}$ to obtain

\begin{equation}
i\left( \partial _{t}+v_{g}\partial _{z}\right) A+\frac{1}{2}v_{g}^{\prime
}\partial _{z}^{2}A-\left[ \frac{\omega }{2\omega _{cp}^{2}}\sum_{e,p}\left[
\omega _{p}^{2}\frac{n_{1}}{n_{0}}\left( 1-\frac{2\omega }{\omega _{c}}%
\right) -\frac{e^{2}\omega ^{2}\omega _{p}^{2}|A|^{2}}{m_{e}^{2}c^{2}\omega
_{c}^{2}}\left( 1-\frac{4\omega }{\omega _{c}}\right) \right] \right] A=0.
\label{eq:vector-expand}
\end{equation}
As we can see, the correction term proportional to $\omega /\omega _{c}$
must be kept above, since there will be an approximate cancellation of terms
proportional to $n_{1}/n_{0}$ and $|A|^{2}$. Correction terms proportional
to the small parameter $1-v_{g}/c$ that have been omitted in Eq. (\ref
{eq:vector-expand}) in general give an additional contribution to Eq. (\ref
{eq:vector-expand}). However, unless we consider the regime where $\omega
<\omega _{p}$ (in which case ion-motion must be included in the
low-frequency dynamics), such terms are of higher order, and hence omitted.
Next combining Eqs. (\ref{eq:cont})-(\ref{eq:Poisson}) we can express the
density perturbations in terms of the ponderomotive force and the potential.
Substituting this expression in Eq. (\ref{eq:vector-general}), using $\omega
\ll \omega _{c}$ and $v_{t}^{2}\equiv T/m_{e}$ $\ll c^{2}$ to simplify the
equation, we find

\begin{equation}
i\left( \partial _{t}+v_{g}\partial _{z}\right) A+\frac{1}{2}v_{g}^{\prime
}\partial _{z}^{2}A+\frac{\omega }{2\omega _{cp}^{2}}\left[ \frac{n_{0i}e^{3}%
}{m_{e}^{2}c^{2}}\phi A-\sum_{e,p}\frac{e^{2}\omega ^{2}\omega _{p}^{2}}{%
m_{e}^{2}c^{2}\omega _{c}^{2}}\left( \frac{v_{t}^{2}}{c^{2}}-\frac{2\omega }{%
\omega _{c}}\right) |A|^{2}A\right] =0.  \label{eq:vector-approx}
\end{equation}
Next, combining Eqs. (\ref{eq:cont})-(\ref{eq:Poisson}) to solve for the
scalar potential, we have

\begin{equation}
\partial _{z}\left( \frac{\partial ^{2}}{\partial t^{2}}%
-(v_{e}^{2}+v_{p}^{2})\frac{\partial ^{2}}{\partial z^{2}}+\omega _{p\mathrm{%
tot}}^{2}\right) \phi =-\frac{\Omega _{i}^{2}}{2m_{e}\omega _{cp}^{2}}%
\sum_{e,p}q\omega _{p}^{2}\frac{\partial |A|^{2}}{\partial z},
\label{eq:pot-first}
\end{equation}
where $\omega _{p\mathrm{tot}}^{2}=\omega _{pe}^{2}+\omega _{pp}^{2}$.
Integrating (\ref{eq:pot-first}), dropping the term proportional to $%
(v_{e}^{2}+v_{p}^{2})/c^{2}$, and using the background charge balance we find

\begin{equation}
\left[ \frac{\partial ^{2}}{\partial t^{2}}+\omega _{p\mathrm{tot}}^{2}%
\right] \phi =\frac{e\Omega _{i}^{4}|A|^{2}}{2m_{e}\omega _{cp}^{2}}.
\label{eq:pot-second}
\end{equation}

The coupled equations (\ref{eq:vector-approx}) and (\ref{eq:pot-second})
describe the nonlinear self-interactions of a high-frequency CPEM pulse
combined with the generation of a low-frequency electrostatic wake field in
a strongly magnetized electron-positron ion plasma. We note that there is no
wake field generation in the absence of the ions.

\section{Conservation laws}

In a coordinate system moving with the group velocity and with suitable
normalizations (i.e. $\xi =\omega _{p\mathrm{tot}}(z-v_{g}t)/c$, etc.), the
above system can be put into the generic form 
\begin{eqnarray}
\mathrm{i}\frac{\partial A}{\partial \tau }+\frac{\partial ^{2}A}{\partial\xi^{2}} 
&=& -A\left( \Phi -\alpha \left| A\right| ^{2}\right)   \label{HF} \\
\frac{\partial ^{2}\Phi }{\partial \xi ^{2}}+\Phi  &=&\left| A\right| ^{2},
\label{LF}
\end{eqnarray}
where $\alpha $ is a constant that determines the relative importance of the
self-nonlinearity, in our case given by 
\begin{equation}
\alpha =\frac{8\pi \omega _{p\mathrm{tot}}^{2}\omega ^{2}}{\Omega _{i}^{6}}%
\left( \frac{\omega _{p\mathrm{tot}}^{2}v_{t}^{2}}{c^{2}}+\frac{2\Omega
_{i}^{2}\omega }{\omega _{cp}}\right) 
\end{equation}
Various aspects of coupled system of this type have been studied by Refs. 
\cite{Scripta2004,Brodin1998,Karpman97-98}. Following the presentation in 
\cite{Scripta2004} we will refer to the low-frequency field $\Phi $ as the
wake-field. The equation system (\ref{HF}) and (\ref{LF}) can be derived
from a variational principle. Introducing the Lagrangian density

\begin{equation}
\mathcal{L}=\frac{1}{2}\left( A^{\ast }\frac{\partial A}{\partial \tau }-A%
\frac{\partial A^{\ast }}{\partial \tau }\right) -\left| \frac{\partial A}{%
\partial \xi }\right| ^{2}+\frac{1}{2}\alpha \left| A\right| ^{4}+\frac{1}{2}%
\left( \frac{\partial \Phi }{\partial \xi }\right) ^{2}-\frac{\Phi ^{2}}{2}%
+|A|^{2}\Phi ,  \label{Lagrangian}
\end{equation}
where the action functional is $\mathcal{A}(\Phi ,A,A^{\ast })=\int \mathcal{%
L}d\xi d\tau ,$ we obtain Eqs. (\ref{HF}) and (\ref{LF}) by varying $\Phi $
and $A^{\ast }$ and minimizing the action as usual. The system posses the
following three conservation laws:

\begin{eqnarray}
\frac{d}{d\tau }\int_{-\infty }^{\infty }\left| A\right| ^{2}d\xi &=&0 ,
\label{quanta} \\
\frac{d}{d\tau }\int_{-\infty }^{\infty }\left( \left| \frac{\partial A}{%
\partial \xi }\right| ^{2}-\frac{\alpha }{2}\left| A\right| ^{4}|-\left|
A\right| ^{2}\Phi \right) d\xi &=&0,  \label{Hamiltonian} \\
\frac{d}{d\tau }\int_{\xi _{-}}^{\xi _{+}}\left( \frac{\partial A}{\partial
\xi }A^{\ast }-\frac{\partial A^{\ast }}{\partial \xi }A\right) &=&\left.
W_{\Phi }\right| _{\xi _{-}}^{\xi _{+}},  \label{energy}
\end{eqnarray}
where $W_{\Phi }=(1/2)[\Phi ^{2}+(\partial \Phi /\partial \xi )^{2}]$ . \
Within the framework of the WKB-approximation, $\int_{-\infty }^{\infty
}\left| A\right| ^{2}d\xi $ is proportional to the number of high-frequency
quanta $N$, and thus Eq. (\ref{quanta}) describes the conservation of $N$.
Using a quantum mechanical analogue, the process of wake field generation
can thus be viewed as a parametric process where a high-frequency quanta
with frequency $\omega $ decays into a low frequency quanta $\Omega $ and
another high-frequency quanta with frequency $\omega -\Omega $, conserving
the total number of high-frequency quanta. Furthermore, Eq. (\ref
{Hamiltonian}) is the conservation equation for the Hamiltonian. Although (%
\ref{energy}) does not look like a conservation law, it actually describes
the conservation of energy. To see this, it is convenient to choose the
positions $\xi _{\pm }$ before and after the pulse passage, respectively.
For this choice we note that the left-hand side of (\ref{energy}) is
proportional to $d/d\tau \int_{-\infty }^{\infty }\kappa |A_{\kappa
}|^{2}d\kappa /\int_{-\infty }^{\infty }|A_{\kappa }|^{2}d\kappa =d\Delta
k/d\tau =(1/v_{g})d\Delta \omega /d\tau $. Furthermore, since $N$ is
conserved, $d\Delta \omega /d\tau $ is proportional to the high-frequency
energy decay rate. Finally, the difference in $W_{\Phi }$ before and after
the pulse passage is proportional to the energy transfer rate to the wake
field, and thus we deduce that Eq. (\ref{energy}) implies energy
conservation.

\section{Summary and conclusions}

To summarize, we have considered the nonlinear interactions between magnetic
field-aligned CPEM waves and electrostatic plasma oscillations (EPOs) in a
pair-ion magnetoplasma. It is found that the EPOs are generated by the
ponderomotive force of the CPEM waves only when the ions in a pair plasma
are present. The present nonlinear wave-wave interactions provide the
possibility of CPEM pulse localization and the generation of wakefields.
Localized EM pulses, in association with wakefields, can be identified in
observations from astrophysical settings as well as from next generation
intense laser-solid density plasma experiments where pairs and ions appear
simultaneously.

\acknowledgments
This research was partially supported by the Swedish Research Council (VR). 

\end{document}